\documentclass[sn-mathphys,Numbered,iicol]{sn-jnl}

\usepackage{graphicx}%
\usepackage{multirow}%
\usepackage{amsmath,amssymb,amsfonts}%
\usepackage{amsthm}%
\usepackage{mathrsfs}%
\usepackage[title]{appendix}%
\usepackage{xcolor}%
\usepackage{textcomp}%
\usepackage{manyfoot}%
\usepackage{booktabs}%
\usepackage{algorithm}%
\usepackage{algorithmicx}%
\usepackage{algpseudocode}%
\usepackage{listings}%
\usepackage{lipsum}
\usepackage{dcolumn}
\usepackage{bm}
\usepackage{csquotes}
\usepackage{enumerate}
\usepackage{commath}
\usepackage{ulem}
\usepackage{braket}
\usepackage{color}
\usepackage{cleveref}
\usepackage{braket}
\usepackage{mwe}
\usepackage{siunitx}
\usepackage{soul}

\DeclareMathOperator{\e}{\mathrm{e}}
  
\begin{document}

\title[Article Title]{Spatial and temporal characteristics of spontaneous parametric down-conversion with varying focal planes of interacting beams}

\author*[1]{\fnm{Richard} \sur{Bernecker}}\email{richard.bernecker@uni-jena.de}

\author[1,2,3]{\fnm{Baghdasar} \sur{Baghdasaryan}}\email{baghdasar.baghdasaryan@uni-jena.de}

\author[1,2,4]{\fnm{Stephan} \sur{Fritzsche}}\email{s.fritzsche@gsi.de}

\affil*[1]{\orgdiv{Theoretisch-Physikalisches Institut}, \orgname{Friedrich-Schiller-Universit\"at Jena}, \orgaddress{\street{Max-Wien-Platz 1}, \city{Jena}, \postcode{ 07743}, \country{Germany}}}

\affil[2]{\orgname{Helmholtz-Institut Jena}, \orgaddress{\street{Fr\"obelstieg}, \city{Jena}, \postcode{07743}, \country{Germany}}}

\affil[3]{\orgname{Fraunhofer Institute for Applied Optics and Precision Engineering IOF}, \orgaddress{\street{Albert-Einstein-Strasse 7}, \city{Jena}, \postcode{07745}, \country{Germany}}}

\affil[4]{\orgname{Abbe Center of Photonics}, \orgaddress{\street{Albert-Einstein-Str. 6}, \city{Jena}, \postcode{07745}, \country{Germany}}}

\abstract{Spontaneous parametric down-conversion (SPDC) is a widely used process to prepare entangled photon pairs. In SPDC, a second-order nonlinear crystal is pumped by a coherent laser beam to generate photon pairs. The photon pairs are usually detected by single-mode fibers (SMF), where only photons in a Gaussian mode can be collected. The collection modes possess typical Gaussian parameters, namely a beam waist and a focal plane position. The collection efficiency of photons highly depends on the choice of both parameters. The exact focal plane position of the pump beam relative to those of the detection modes is difficult to determine in a real experiment. Usually, theoretical and experimental studies assume that the focal plane positions of the pump and the generated beams are positioned in the center of the crystal. The displacement of beam focal planes can lead to deviations from expected results and the coupling efficiency into SMF can decrease. In this study, we theoretically examine variable positions of focal planes in the Laguerre-Gaussian basis, a popular experimental modal decomposition of the spatial biphoton state. We explore how the choice of focal plane positions affects the spatial and temporal properties and the purity of the photon pairs. We present SPDC setups where precise knowledge of the focal plane position is essential and scenarios where focal plane displacements have negligible impact on experimental outcomes.}


\maketitle

\section{Introduction}

Quantum-based technologies are increasingly explored and integrated into today's applications. In this context, quantum entanglement is an inseparable part of quantum communication protocols \cite{wengerowsky2020passively, yin2020entanglement}. The process of spontaneous parametric down-conversion (SPDC) is the most reputable source of photonic entanglement \cite{Friis2019, krenn2014generation} and provides an experimental platform for fundamental quantum science \cite{anwar2021entangled}. 

In SPDC, a nonlinear crystal is illuminated with a strong, high-energetic laser field called pump beam. Photon pairs with lower energies, also known as signal and idler photons, are subsequently down-converted. The generated signal and idler photons fulfill the energy $\omega_p = \omega_s + \omega_i$ and the momentum $\bm{k}_p = \bm{k}_s + \bm{k}_i$ conservations. The momentum conservation, also known as the phase-matching (PM) condition, ensures constructive interference between the three interacting beams and inherently determines the spectral and spatial properties of signal and idler photons. Spectrally and spatially engineered photons are important ingredients in current research on quantum information \cite{QiProcessing2019, caspani2017integrated, moreau2019imaging}, quantum computing \cite{zhong2020quantum}, and quantum communication \cite{sansa2022visible, schimpf2021entanglement}.
Additionally to the PM, the pump beam properties also have a large impact on the spectral and spatial properties of signal and idler photons \cite{JPTorres1, francesconi2021anyonic, Kovlakov2018}. The pump characteristics include the beam width, the focal plane position relative to the crystal, and its spatial and temporal degrees of freedom (DOFs) \cite{pires2011, buse2015photons}.

\begin{figure*}
    \centering
    \includegraphics[width= 0.99\textwidth]{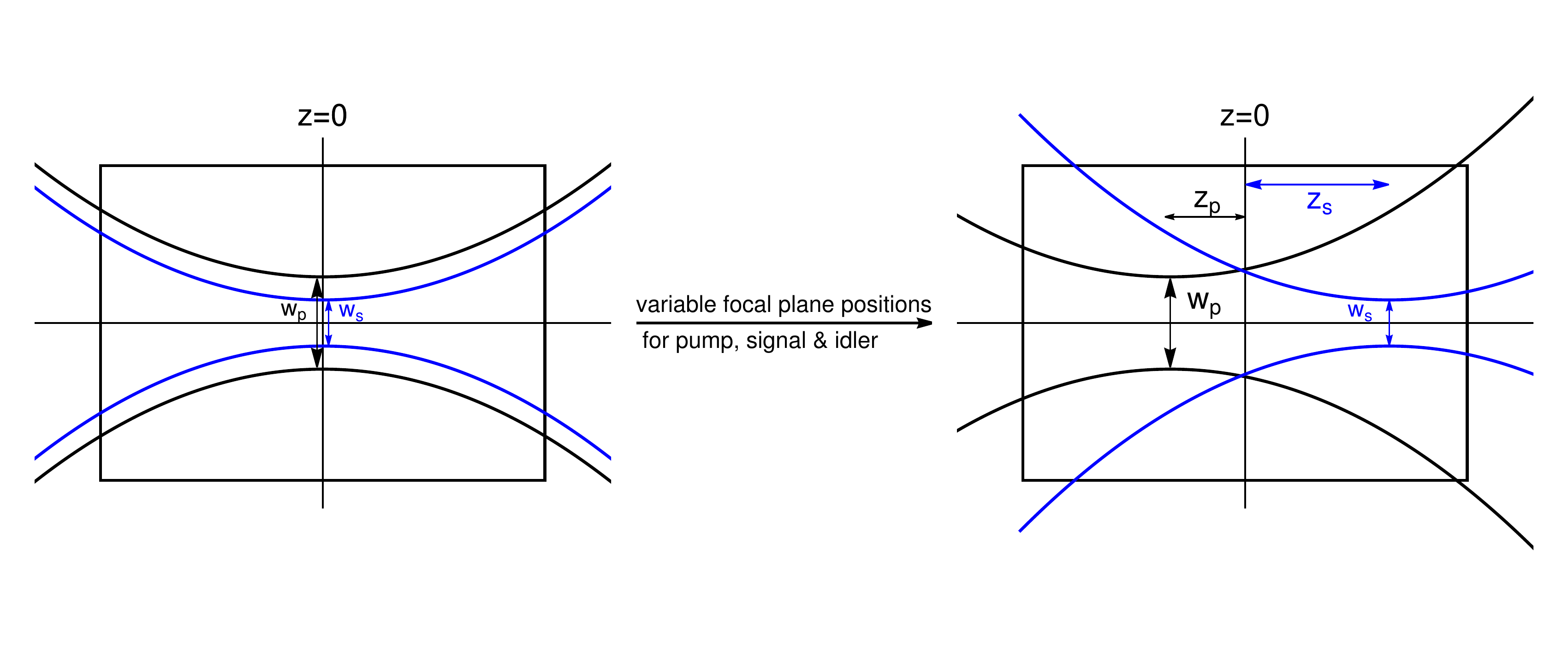}
    \vspace{-1cm}
    \caption{Schematic paths of pump and signal beam in a non-linear crystal. For the sake of simplicity, only the signal beam is shown, the idler beam can be similarly imagined. The beams are described as Gaussian beams with beam widths $w_p$ and $w_s$. Most calculations assume that the focal planes of the pump, signal, and idler lie in the center of the crystal $z=0$ as shown in the left picture. On the contrary, we allow in the right picture that the pump, signal, and idler focal planes are not fixed. The parameters for focal plane shifts along the propagation axis are $z_p$ for the pump beam and $z_s, z_i$ for the generated beams.}
    \label{figure1}
\end{figure*}

Besides the generation of photon pairs, optimal experimental verification is an essential part of quantum fundamental research too. Usually, the spatial shape of signal and idler photons are detected by multi-plane light conversion (MPLC) \cite{Morizur:10,fontaine2019laguerre}, where an arbitrary spatial mode is projected to a Gaussian mode \cite{sevillakaipalathetc2022}, in order to couple it into a single-mode fiber (SMF) \cite{zhang14radial, bouchard2018measuring}. The coupling efficiency into SMF depends on the beforehand chosen focal planes and beams widths of the pump, signal, and idler modes.  Optimizing the coupling efficiency of collecting a photon pair in fundamental Gaussian modes (FGMs) is a particularly interesting aspect from an experimental perspective \cite{guerreiro2013high, bennink2010, bennink2011, jabir2017robust}. We distinguish between the single-mode coupling efficiency for a certain frequency (within a narrowband filter bandwidth) \cite{smirr2013optimal}, and the spectral brightness, which pertains to the maximal collection probability on a broadband of frequencies \cite{steinlechner2012high, steinlechner2014efficient}.

In the past, there were several theoretical and experimental approaches to modify the pump or the detection scheme of the generated modes with linear optical systems \cite{Klyshko1994, Pittman95, Pittman96, Monken98_1, Rubin96}, in order to improve the pair collection efficiency. The variation of the pump intensity or the pump beam width has also been explored to optimize the photon pair collection efficiency or transverse spatial correlations \cite{Monken1998_2, suzer2008, Giuseppe2003effective}. In this regard, the manipulation of the purity between the signal and idler photons via pump focusing was shown \cite{Woerdmann2005, Bennink2006}. Moreover, it has been proposed in Refs. \cite{Karanetal, WALBORN2010} how to consider the change of the angular spectrum of the pump at different positions beyond the crystal center. In analogy to the change of the pump focal plane position, the position of the nonlinear crystal has been varied to control the time delay between signal and idler photons and the coincidence rate \cite{buse2015photons}.

All these considerations were primarily concerned with optimal pump focusing or the right choice of optical elements in beam paths, in order to achieve enhanced photon collection efficiency in SPDC. In this work, we theoretically consider variable focal planes for pump, signal, and idler explicitly and describe their impact on experimentally measurable quantities such as the coupling efficiency into SMF, spatial and temporal correlations of generated photons and the spectral purity between down-converted photons. We describe the biphoton state in the Laguerre-Gaussian (LG) basis, an experimentally readily accessible basis approximating the Schmidt decomposition \cite{miatto2012spatial} with great practical relevance in experiments that leverage fiber coupling or optical orbital angular momentum (OAM) \cite{valencia2021entangled, yang2022generation, krenn2017orbital}. In this regard, we will compare scenarios of focal planes fixed at different positions and investigate if the measurement probability of signal and idler photons remain unaffected. We discuss setup conditions with noteworthy influence on the spectral brightness and the coupling efficiency. The alignment of focal plane positions in these scenarios will require more careful effort in order to increase the coupling efficiency. We also contemplate scenarios where the precise position of focal planes becomes insignificant. These findings are of particular interest in terms of enhancing the efficiency of SPDC experiments utilizing LG modes.

\section{Theory}

We start our investigation with the theoretical description of the SPDC process. Our group published a paper on the characterization of spatio-temporal DOFs in SPDC, where a general expression for the SPDC-state, also known as biphoton state, has been derived \cite{baghi2022} and later verified experimentally \cite{sevillakaipalathetc2022}. First, we briefly recap the derivation of the expression from \cite{baghi2022}, where the focal planes of the pump, signal, and idler beams are assumed to be at $z=0$, i.e in the crystal center. This assumption, as shown in Fig. \ref{figure1} on the left, is common in theoretical as well as experimental studies. Next, we extend the expression to consider variable focal planes for the pump, signal, and idler beams, which is illustrated on the right side of Fig. \ref{figure1}. 

\subsection{Biphoton state of SPDC decomposed in Laguerre-Gaussian Basis}

The common expression of the general biphoton state in the wave vector representation of the interacting beams is \cite{WALBORN2010}
\begin{align}
     \ket{\Psi}_{\text{SPDC}} = \iint & d\bm{q}_s \: d\bm{q}_i \: d\omega_s \: d\omega_i \: \Phi(\bm{q}_s,\bm{q}_i,\omega_s,\omega_i) \nonumber \\ &\times \hat{a}_s^{\dagger}(\bm{q}_s,\omega_s) \: \hat{a}_i^{\dagger}(\bm{q}_i,\omega_i) \ket{\text{vac}},
     \label{commonexpr}
\end{align}
where we consider the paraxial approximation by the separation into longitudinal and transversal components of the wave vector $\bm{k} = \bm{q} + k_z (\omega)  \: \bm{z}$, where $z$ is the propagation direction of the pump beam. The paraxial approximation is valid in most experimental SPDC setups since typical optical apparatuses support only paraxial rays about the central axis. The biphoton state \eqref{commonexpr} is an integral over all possible transverse wave vectors $\bm{q}_{s,i}$ and frequencies $\omega_{s,i}$ of a signal and idler pair that is created from the vacuum state $\ket{\text{vac}}$ by creation operators $\hat{a}_{s,i}^{\dagger}(\bm{q}_{s,i},\omega_{s,i})$ of signal and idler photons, respectively.

The so-called biphoton mode function $\Phi(\bm{q}_s,\bm{q}_i,\omega_s,\omega_i)$ encodes the coupling between the wave vectors of the pump, signal, and idler beams \cite{Karanetal}:
\begin{align*}
    \Phi(\bm{q}_s,\bm{q}_i,\omega_s, & \omega_i) = N \: V_p(\bm{q}_s+\bm{q}_i) \: S_p(\omega_s + \omega_i) \nonumber \\ &\times \int_{-L/2}^{L/2} dz \: \e^{i (k_{p,z} - k_{s,z} - k_{i,z}) z}, \nonumber
\end{align*}
where $N$ is the normalization constant, $V_p$ is the spatial distribution of the pump beam, whereas $ S_p$ characterizes the spectral DOF of the pump. The integral over the propagation direction $z$ describes the phase mismatch ${\Delta k_z = k_{p,z} - k_{s,z} - k_{i,z}}$ in the $z$ direction. The exact expression of $\Delta k_z$ depends on the features of the crystal and the geometry between the interacting beams and the crystal.

Since discrete modes are easier to detect and manipulate in experiments \cite{Eckstein:11, bolduc2013exact} than continuous modes, the continuous transverse spatial variables in \eqref{commonexpr} are often discretized by a set of optical modes. Furthermore, an appropriate choice of the set can reduce the dimensionality of a state. In Ref. \cite{baghi2022}, Laguerre-Gaussian (LG) modes have been used as a basis for the description of the spatial distribution of the down-converted photons. This choice is reasonable since LG modes carry well-defined projection of orbital angular momentum (OAM) \cite{fickler2012quantum}, which is conserved in collinear SPDC \cite{mair2001entanglement,Yao, MiattoBarnettYao}. The biphoton state decomposed in the LG basis $\ket{p,\ell,\omega}=\int  d\bm{q}\, \mathrm{LG}_{p}^{\ell}(\bm{q})\,  \hat{a}^{\dagger}(\bm{q},\omega) \ket{\text{vac}}$  reads then
\begin{align*}
    \ket{\Psi}_{\text{SPDC}} = \iint & d\omega_s \: d\omega_i \sum_{p_s,\ell_s}  \sum_{p_i,\ell_i} C_{p_s,p_i}^{\ell_s,\ell_i}(\omega_s,\omega_i) \nonumber \\  & \times \ket{p_s,\ell_s,\omega_s} \ket{p_i,\ell_i, \omega_i},
\end{align*}
where the overlap amplitudes $C_{p_s,p_i}^{\ell_s,\ell_i}$ of the LG decomposition are frequency-dependent. The probability to find signal and idler photons in spatial modes $(p_s|\ell_s)$ and $(p_i|\ell_i)$ at frequencies $\omega_s$ and $\omega_i$ is given by $P_{p_s,p_i}^{\ell_s,\ell_i}(\omega_s,\omega_i) = |C_{p_s,p_i}^{\ell_s,\ell_i}(\omega_s,\omega_i)|^2$. We can also call this quantity the single-mode \textit{coupling efficiency}. On the other hand, the maximal value of $P_{p_s,p_i}^{\ell_s,\ell_i}(\omega_s,\omega_i)$ over all possible energies $\omega_s$ and $\omega_i$ is called the \textit{spectral brightness}.

The following assumptions and approximations have been applied in Ref. \cite{baghi2022}, in order to derive a comprehensive expression for $C_{p_s,p_i}^{\ell_s,\ell_i}(\omega_s,\omega_i)$:

\begin{itemize}
    \item Pump, signal, and idler beams are focused in the middle of the crystal.
   
    \item A small deviation $\Omega$ of generated frequencies from the central frequency $\omega_0$ has been assumed, i.e. $\Omega\ll\omega_0$, so that we can write $\omega = \omega_0 + \Omega$. The central frequencies fulfill the energy conservation $\omega_{p,0} = \omega_{s,0} + \omega_{i,0}$.
    
    \item In the paraxial regime, where  $|\bm{q}| \ll k$, the so-called Fresnel approximation can be applied on $k_z = \sqrt{k^2 - \abs{\bm{q}}^ 2}$:
        \begin{align}
            k_z & =k(\Omega)\sqrt{1-\frac{|\bm{q}|^2}{k(\Omega)^2}} \nonumber \\
            & \approx k+\frac{\Omega}{u}+\frac{G\Omega^2}{2}-\frac{|\bm{q}|^2}{2k},
            \label{fresnelapprox2}
        \end{align}
        with the group velocity $u=1/(\partial k/\partial \Omega)$ and the group velocity dispersion $G=\partial/\partial \Omega \,(1/u)$ at the corresponding central frequency. 
     \item  Momentum conservation for the central frequencies $\Delta k = k_p - k_s - k_i = 0$ is assumed. When a periodically poled crystal with poling period $\Lambda$ along the crystal axis is used, this is achieved by $\Delta k = k_p - k_s - k_i - \frac{2\pi}{\Lambda}  = 0$.
\end{itemize}

We shall expand now the formalism from Ref. \cite{baghi2022}, to support variable positions of the focal planes for the pump, signal, and idler beams.

\subsection{Shift of focal plane positions}

We briefly recap the angular spectrum propagation of beams. Mathematically, electromagnetic field distributions can be described by a propagator factor obtained via the angular spectrum representation in momentum space. This is a well-investigated formalism, with the following main key ideas. We can choose the $z$-direction as the propagation axis and write the Fourier transformation of an arbitrary field $\bm{E}$ in the transverse $x$-$y$-plane of a fixed point $z=\text{const.}$ as
\begin{equation*}
    \bm{\tilde{E}}(\bm{q};z) \propto \iint_{-\infty}^{\infty} dx \: dy \: \bm{E}(x,y,z) \e^{-i k_x x} \: \e^{-i   k_y y}.
\end{equation*}
Here are $\bm{q}=(k_x,k_y)$ the transverse wave vector components. The amplitude in momentum-space $\bm{\tilde{E}}(\bm{q};z)$ can also be used for the inverse Fourier transform for the field in real space \hspace{-2cm}
\begin{equation*}
    \bm{E}(x,y,z) \propto \iint_{-\infty}^{\infty} dk_x \: dk_y \: \bm{\tilde{E}}(\bm{q};z) \e^{i  k_x x} \: \e^{i   k_y y}.
    \label{inverse FT}
\end{equation*}
When we split the field in a spatial and time-dependent part $\bm{E} = \bm{E}(x,y,z) \e^{-i \omega t} +\text{c.c}$ and also write 
$|\bm{k}|= k = \frac{\omega^2 n^2}{c^2}$, the Helmholtz equation reads as
\begin{equation*}
    \left( \bm{\nabla}^2 + k^2 \right) \bm{E}(x,y,z) = 0.
\end{equation*}
We insert $\bm{E}(x,y,z)$ into the Helmholtz equation and arrive at a differential equation for the spatial evolution:
\begin{equation*}
    (\partial_z^2 + k^2 - k_x^2 -k_y^2 )  \:\bm{\tilde{E}}(\bm{q};z) = 0.
\end{equation*}
When setting $ k_z =\sqrt{k^2 - k_x^2 -k_y^2}$, the solution of the angular spectrum of an electric field evolving along the z-axis can be written as
\begin{equation}
    \bm{\tilde{E}}(\bm{q};z)= \bm{\tilde{E}}(\bm{q};0) \:  e^{\pm i k_z z}
    \label{displacement}
\end{equation}
(see also Refs. \cite{WALBORN2010, Karanetal}). The positive signs in the exponential indicate a wave propagation into $z > 0$, while the negative sign describes a wave propagating into the region $z < 0$.

We apply Eq. \eqref{displacement} to the pump, signal, and idler beams. The PM function renewed with the pump, signal, and idler focused at positions $z_p$, $z_s$, and $z_i$ respectively (see Fig. \ref{figure1}), is now
\begin{align*}
   \Phi(&\bm{q}_s,\bm{q}_i,\omega_s, \omega_i) = V_p(\bm{q}_s+\bm{q}_i) \: S_p(\omega_s + \omega_i) \nonumber \\  & \times \int_{-L/2}^{L/2} dz \: \e^{i \left[ k_{p,z} (z + z_p) - k_{s,z} (z+ z_s) - k_{i,z} (z+ z_i)  \right]}.
   \label{modified}   
\end{align*} 
Note that we continue to consider paraxial beams for pump, signal, and idler and use the Fresnel approximation from Eq. \eqref{fresnelapprox2} for $k_z$. 

To specify our setup, we assume a Gaussian envelope of pulse duration $T_0$ for the spectral part of the pump. Due to $\omega_p-\omega_{p,0} = \Omega_p = \Omega_s + \Omega_i$ we can write
\begin{equation}
    S_p(\Omega_p)= \frac{T_0}{\sqrt{\pi}} \exp{ \biggl(-\frac{T_0^2}{4} \: (\Omega_s+\Omega_i)^2 \biggl)}.
    \label{gaussianenvelope}
\end{equation}
The spatial distribution of the pump beam is also described as a Gaussian with beam width $w_p$, 
\begin{equation*}
   \mathrm{V}(\bm{q}_\mathrm{s}+\bm{q}_\mathrm{i}) \: = \: \frac{w_p}{\sqrt{2 \pi}}\:\exp{\biggl(-\frac{w_p^2}{4}|\bm{q}_\mathrm{s}+\bm{q}_\mathrm{i}|^2\biggr).}
\end{equation*}
This reduces and simplifies the expression from Ref.\cite{baghi2022} enormously. The final formula reads then
\begin{strip}
\begin{align} 
      C_{p_s,p_i}^{\ell_s,\ell_i} &
      \propto \delta_{\ell_s,-\ell_i}  \: \exp{ \biggl(-\frac{T_0^2}{4} \: (\Omega_s+\Omega_i)^2 \biggl)} \sum_{s=0}^{p_s}\sum_{i=0}^{p_i} \: (T_s^{p_s,\ell_s})^* \:(T_i^{p_i,\ell_i})^* \: \Gamma[h]\:\Gamma[b]  
     \nonumber\\ \times 
     \int_{-L/2}^{L/2}dz\: & \exp{\biggl[i z\biggl(\frac{\Omega_s + \Omega_i}{u_p}-\frac{\Omega_s}{u_s} - \frac{\Omega_i}{u_i} + \frac{ G_p}{2} (\Omega_s+\Omega_i)^2- \frac{G_s}{2} \Omega_s^2 -\frac{G_i}{2} \Omega_i^2 \biggl)\biggl]} \frac{D^{\ell_i}} {H^{h}\: B^{b}}\: {_2}{\Tilde{F}}_1\biggl[h,b, 1+\ell_i,\frac{D^2}{H \,B}\biggl]
\label{FullExpression}
\end{align} 
\end{strip}
with the abbreviations
\begin{eqnarray*}
  h &=& 1+s+\frac{1}{2} \: (\ell_i+\abs{\ell_s}), \\[0.1cm]
  b &=& 1+i+\frac{1}{2} \: (\ell_i+\abs{\ell_i}),\\[0.1cm]
  D& = & -\frac{ w_p^2}{4}-\frac{i}{2k_p}(z+z_p), \\[0.1cm]
  H &=& \frac{w_p^2}{4}+\frac{w_s^2}{4}- \frac{i}{2} \Bigl[ \frac{(z+z_p)}{k_p} - \frac{(z+z_s)}{k_s} \Bigl], \\[0.1cm]  
  B &=& \frac{w_p^2}{4}+\frac{w_i^2}{4}-\frac{i}{2} \Bigl[\frac{(z+z_p)}{k_p} - \frac{(z+z_i)}{k_i} \Bigl], \\[0.1cm]
  \hspace{-0.5cm}
   T^{p,\ell}_k &=& \frac{(-1)^{p+k}(i)^{\ell}}{(p - k)! \: (|\ell| + k)!k!} \: \sqrt{\frac{p!\,(p+|\ell|)!}{\pi}} \biggr(\frac{ w}{\sqrt{2}}\biggl)^{2k+|\ell|+1}
\end{eqnarray*}
and the \textit{regularized} \textit{hypergeometric} function ${_2}{\Tilde{F}}_1$ \cite{Hypergeometric2F1}. The collecting widths of the generated signal and idler modes are denoted by $w_s$ and $w_i$. The expression \eqref{FullExpression} gives full insight into the spatial distribution of the biphoton state decomposed in LG modes and also into the spatio-temporal coupling in SPDC \cite{Osorio_2008,gatti12dim}. Note that the overlap amplitudes $C_{p_s,p_i}^{\ell_s,\ell_i}$ from Eq. \eqref{FullExpression} depend only on $|\ell|$, where $\ell = \ell_s = - \ell_i$ (see Ref. \cite{sevillakaipalathetc2022}), but we keep the notation of Eq. \eqref{FullExpression} for a proper illustration of our results.

\section{Results and discussion}

\subsection{Justifying the choice $z_s=z_i$}

In this section, we study the impact of $z_p$, $z_s$, and $z_i$ on the probability to detect the signal and idler photons in FGMs, or in other words, the efficiency of direct coupling of generated photons into SMF. In the following sections, unless otherwise stated, we assume spectrally a continuous-wave pump such that $S_p(\omega_s+\omega_i) = \delta(\Omega_s+\Omega_i)$ which leads to the same amount of deviation from the center frequency for signal and idler photons, $\Omega_s=-\Omega_i$.
\begin{figure}
    \centering
    \includegraphics[width=.495 \textwidth]{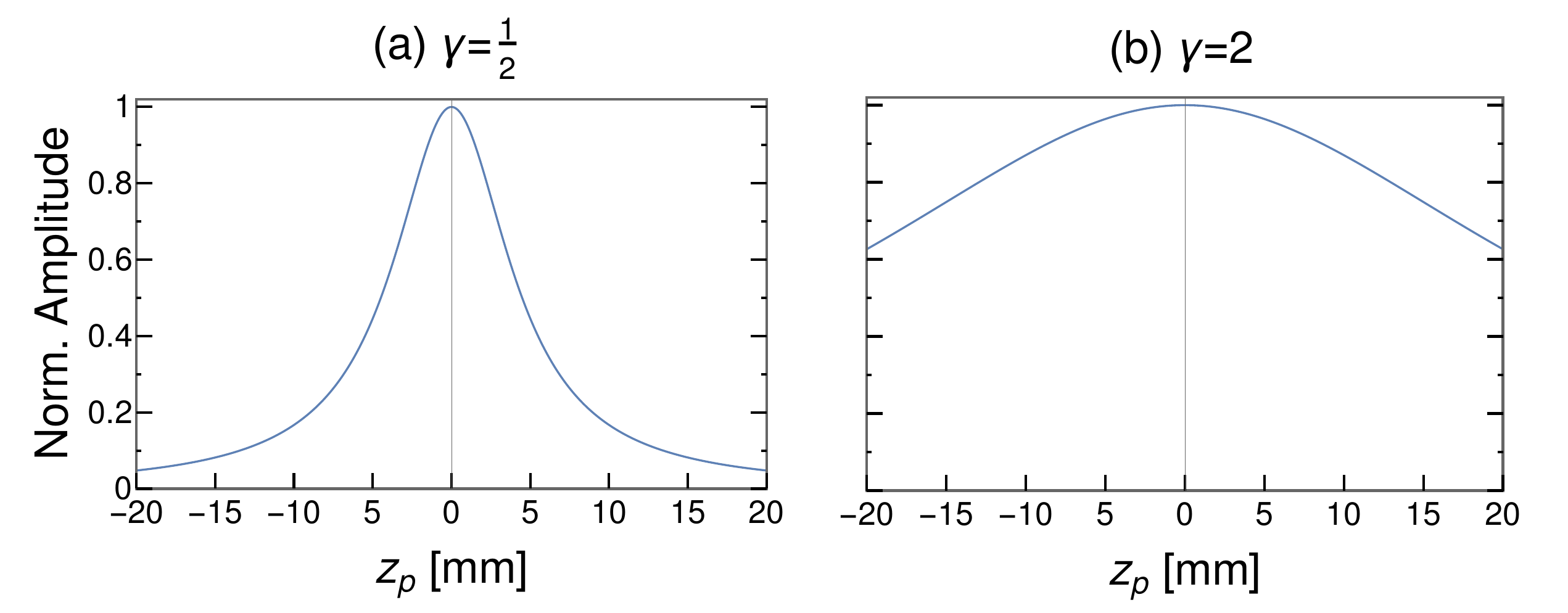}
    \caption{Normalized amplitude of the single-mode coupling for $\lambda_s=\SI{810}{\nano \meter}$ depending on the pump focal position $z_p$ for a crystal with $L=\SI{30}{\milli \meter}$ centered at $z=\SI{0}{\milli \meter}$. The focal plane positions of signal and idler are set at $z_s=z_i=\SI{0}{\milli \meter}$. Two scenarios are examined: (a) {$\gamma = \frac{\SI{10}{\micro \meter}}{\SI{20}{\micro \meter}}$} and (b) $\gamma = \frac{\SI{40}{\micro \meter}}{\SI{20}{\micro \meter}}$. The full width at half maximum is independent of the crystal length but determined by the beam width ratio $\gamma$. As $\gamma$ increases, the influence of the pump focal plane shifts on the normalized amplitude decrease.}
    \label{figure2}
\end{figure}

In general, we identify that the coupling efficiency into SMF decreases, when the pump focal plane is displaced from the crystal center (see Fig. \ref{figure2}). The amplitude becomes more robust to the change of $z_p$ if the beam width ratio $\gamma = \frac{w_p}{w_{s}}$ increases ($w_s=w_i$ is assumed).
\begin{figure*}
    \centering
    \includegraphics[width= 0.95\textwidth]{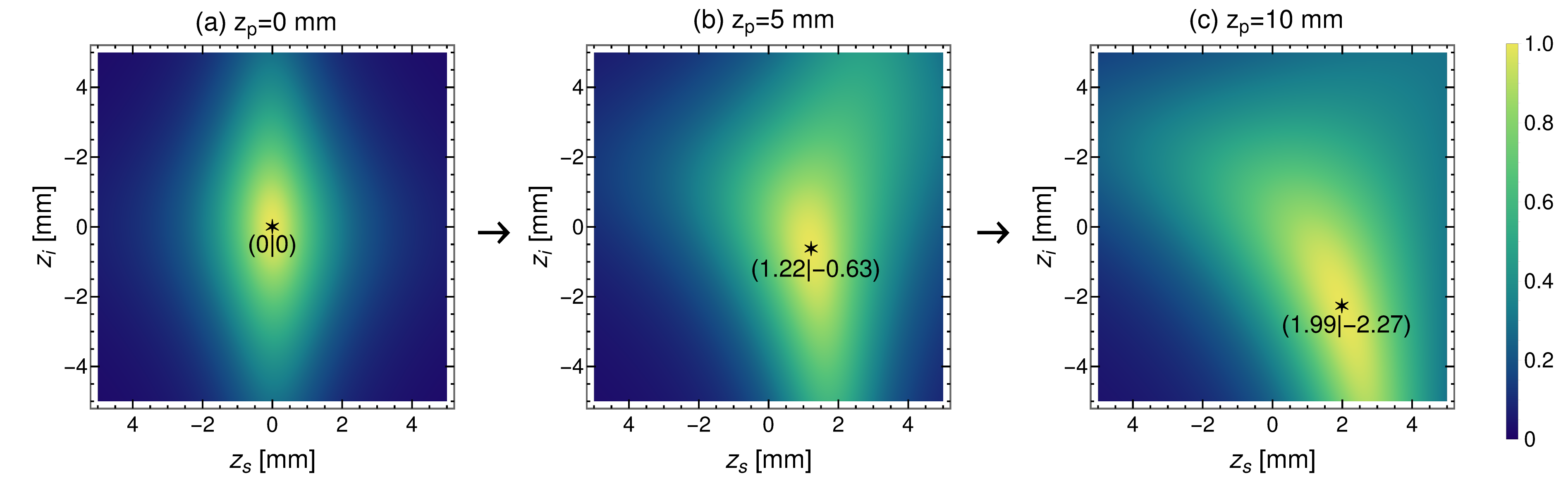}
    \caption{The single-mode coupling efficiency for FGMs in dependence of signal and idler focal plane shits $z_s,z_i$ for different pump focal plane shifts $z_p$. The setup parameters are $L=\SI{1}{\milli \meter}, w_p=w_s=\SI{10}{\micro \meter}, w_i=\SI{20}{\micro \meter}$ and $\lambda_s = \SI{810}{\nano \meter}$. The black star refers to the pair ($z_s|z_i$) that maximizes the coupling efficiency. 
    (a) The pump has its focal plane in the center of the crystal, so optimal coupling would imply that the focal planes of the signal and idler move in opposite directions. (b) and (c) If $z_p \neq \SI{0}{\milli \meter}$, the optimum amplitudes are reached for signal and idler focal planes shifted from the center. These findings are in line with the advanced-wave picture.}
    \label{figure3}
\end{figure*}
In order to increase the amplitude for a given pump focal plane shift $z_p$, we should adjust the signal and idler focal plane positions. The optimal choice of a pair $z_s$ and $z_i$ strongly depends on the beam width of collection modes, $w_s$ and $w_i$. We distinguish two scenarios of equal $w_s =w_i$ or unequal beam widths. 

Fig. \ref{figure3} represents the coupling efficiency for signal photons filtered at $\lambda_s=\SI{810}{\nano \meter}$ for unequal beam widths $w_s \neq w_i$ as a function of $z_s$ and $z_i$. The calculations have been carried out for three different pump focal plane positions $z_p$. The star corresponds to the ($z_s|z_i$) combination that optimizes the single-mode coupling efficiency. If $z_p$ and for instance, $z_s$ are fixed in the experimental setup, the efficiency strongly depends on the focal plane position of the corresponding partner $z_i$. An edge case is visualized in Fig. \ref{figure3} (a), where the pump is fixed in the crystal center. The maximum is attained when the signal focus plane and the idler focus plane displacements are equal in magnitude but point in opposite directions, i.e. $z_i=-z_s$. The initial dependence of the amplitude on $z_s$ and $z_i$ from (a) is more distorted and moves away from the plot center the larger $z_p$ is. Moreover, the maximum amplitudes lay not on the diagonal $z_s = z_i$ and move further away for increasing $z_p$. 

\begin{figure}
    \centering
    \includegraphics[width=.48\textwidth]{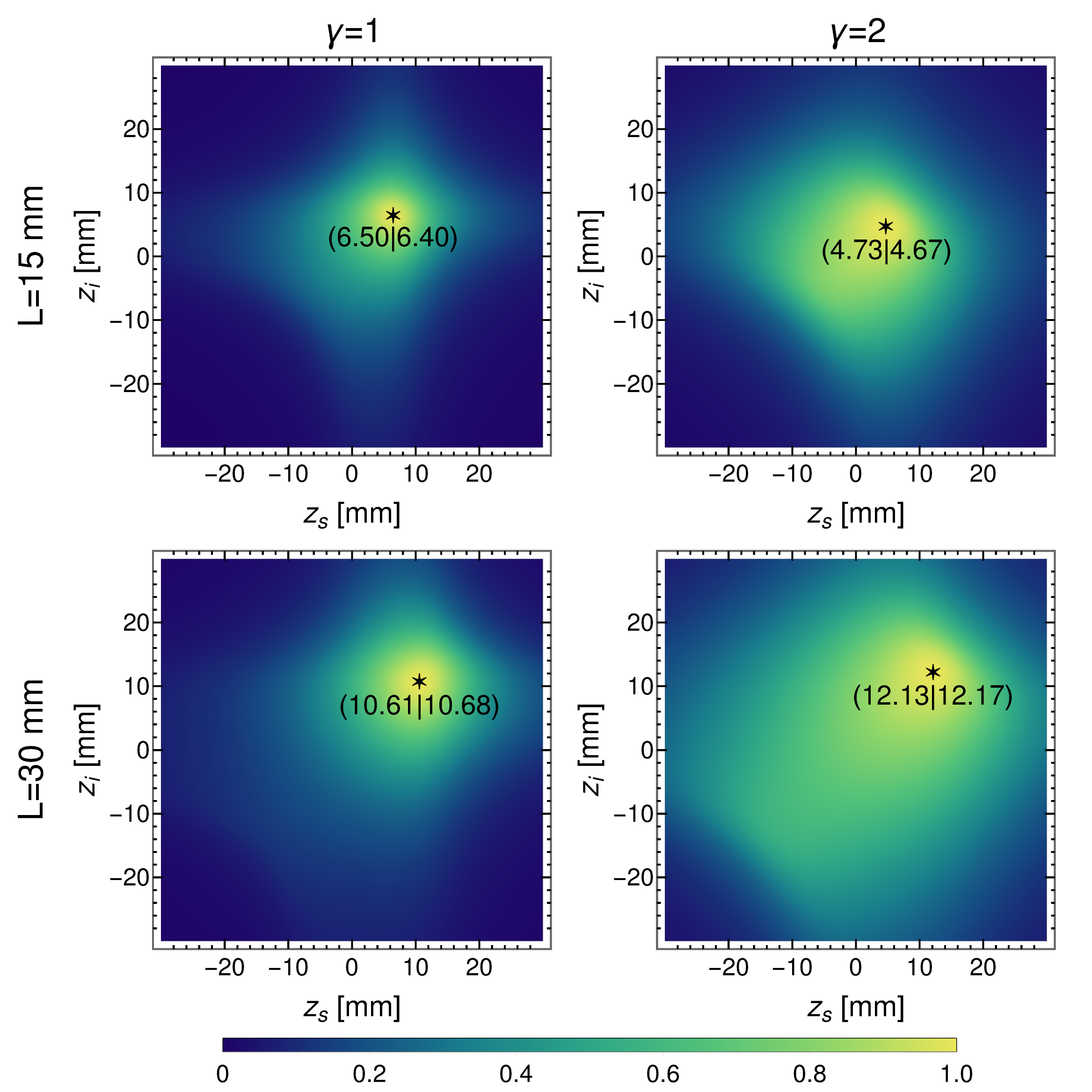}
    \caption{Like in Fig. \ref{figure3} The single-mode coupling efficiency for FGMs is shown in dependence of signal and idler focal plane shifts for a fixed pump focal plane position $z_p=\SI{10}{\milli \meter}$. We considered $w_p=\SI{20}{\micro \meter}, \: \SI{40}{\micro \meter}, w_s=w_i=\SI{20}{\micro \meter}, \lambda_s = \SI{810}{\nano \meter}$. The black star refers to the pair ($z_s|z_i$) that maximizes the coupling efficiency.  The high amplitude areas widen and the exact focal plane position of signal and idler is less relevant for thicker crystals or higher beam width ratios.}
    \label{figure4}
\end{figure}

However, in experiments $w_s = w_i$ is more common and we observe more symmetric dependencies of the single mode coupling efficiency on $z_s$ and $z_i$. Fig. \ref{figure4} shows the same as Fig. \ref{figure3} but for a constant focal plane position $z_p = \SI{10}{\milli \meter}$ and the same beam width for signal and idler $w_s=w_i$. We distinguish between two different crystal lengths $L$ and beam width ratios $\gamma$. Since we choose $z_p \neq \SI{0}{\milli \meter}$, the area of high efficiency lies not around $z_s=z_i=\SI{0}{\milli \meter}$. This area resembles a circle laying on the diagonal $z_s = z_i$. The higher the length or the beam width ratio, the bigger the red circle which means a larger range of $z_s, z_i$ where the amplitude is constant. This enables fixing the focal planes of signal and idler at the same location.

\begin{figure*}
    \centering
    \includegraphics[width= 0.95\textwidth]{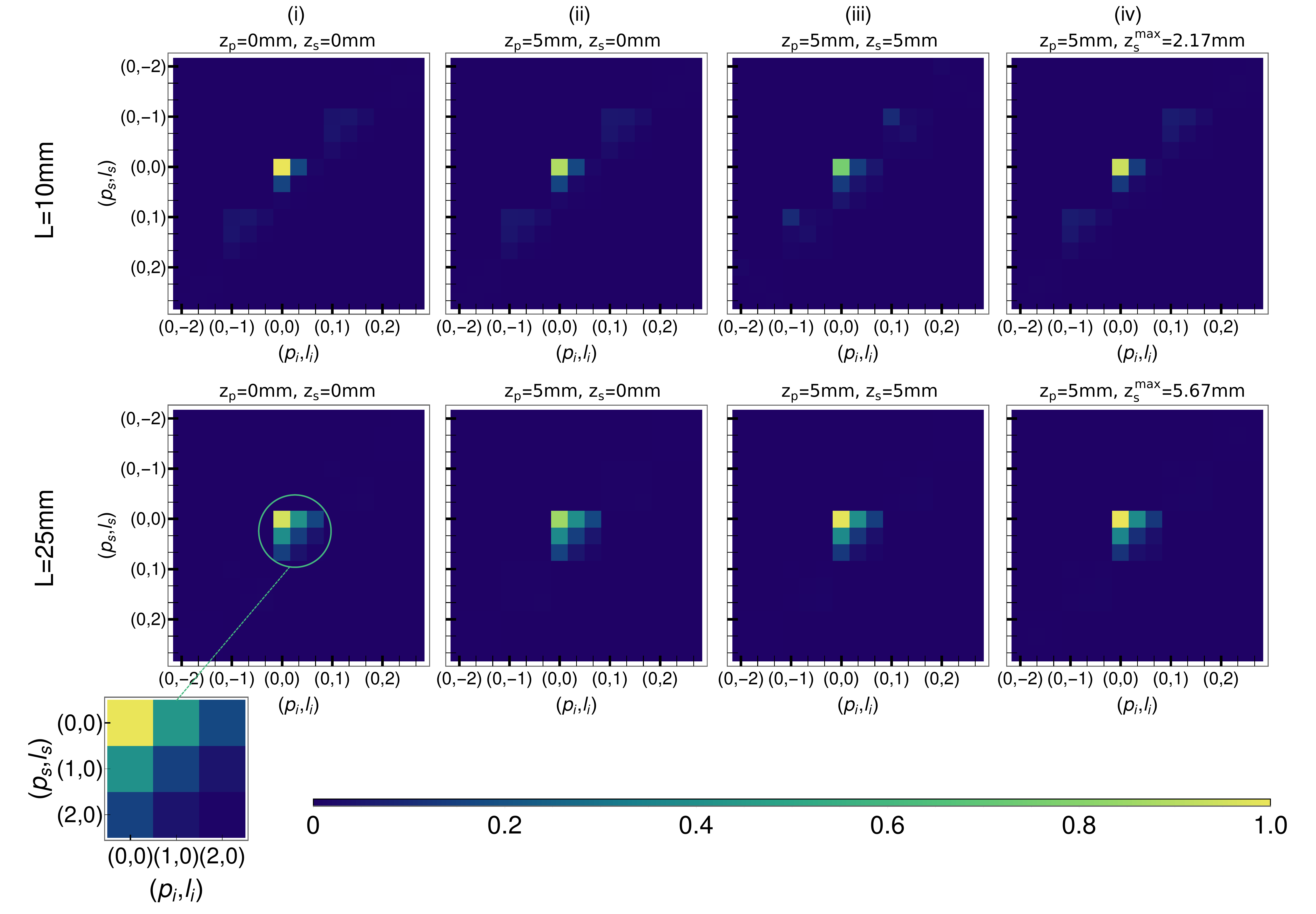}
    \caption{Mode distribution of the biphoton state in LG basis filtered at $\lambda_{s}=\SI{810}{\nano \meter}$ for four different arrangements of focal plane shifts $z_p$ and $z_s=z_i$: (i) pump, signal, and idler focal planes are assumed to be at $z=\SI{0}{\milli \meter}$, (ii) only the pump beam is shifted (by $\SI{5}{\milli \meter}$), but signal and idler remain in the center of the crystal, (iii) pump, signal, and idler are shifted by the same amount of $\SI{5}{\milli \meter}$, (iv) value $z_s=z_i$ that maximizes the coincidence amplitude for the given pump shift $z_p$. The upper show a crystal with length $L=\SI{10}{\milli \meter}$, the lower row represents $L=\SI{25}{\milli \meter}$. The mode numbers $(p_s|\ell_s)$ for signal and  $(p_i|\ell_i)$ for idler run over $p=0,1,2$ and $\ell = -2, \dots ,2$. The thick bars mark p=0 and the two following thin bars p=1,2. Each row is normalized by its maximum (corresponding to crystal length). If the pump is not focused in the center, the signal and idler beams should be shifted as well, to maximize the coupling efficiency.}
    \label{figure5}
\end{figure*}

Our results show that for a fixed frequency when considering the single-mode coupling efficiency, $z_p=\SI{0}{\milli \meter}$ not always implies $z_s=z_i=\SI{0}{\milli \meter}$ for the highest amplitude. This is shown in more detail in chapter \ref{Influence of focal plane shits on the temporal spectrum}. If the total frequency spectrum on contrary is considered, $z_p=\SI{0}{\milli \meter}$ always implies $z_s=z_i=\SI{0}{\milli \meter}$ for maximal "spectral brightness". The focal planes of all beams should lie in the center of the crystal.  

Our findings that focal plane shifts of the pump, signal, or idler beam from the crystal center affect the efficiency and have to compensate each other is consistent with the \textit{advanced-wave picture} (AWP) \cite{Klyshko1994,WALBORN2010, Arruda2018, Ribeiro2020, Aspden2014}, which provides a classical analog to understand biphoton coincidence experiments. 

To summarize, the choice of pump, signal, and idler focal plane position can greatly influence the coupling efficiency. When choosing equal collecting widths for signal and idler $w_s=w_i$, it is sufficient to assume signal and idler focal plane positions at the same spot  $z_s = z_i$ for maximum efficiency. 

\subsection{Spatial and temporal characteristics for $z_s=z_i$} 

\begin{figure*}
\centering
    \includegraphics[width= 0.95 \textwidth]{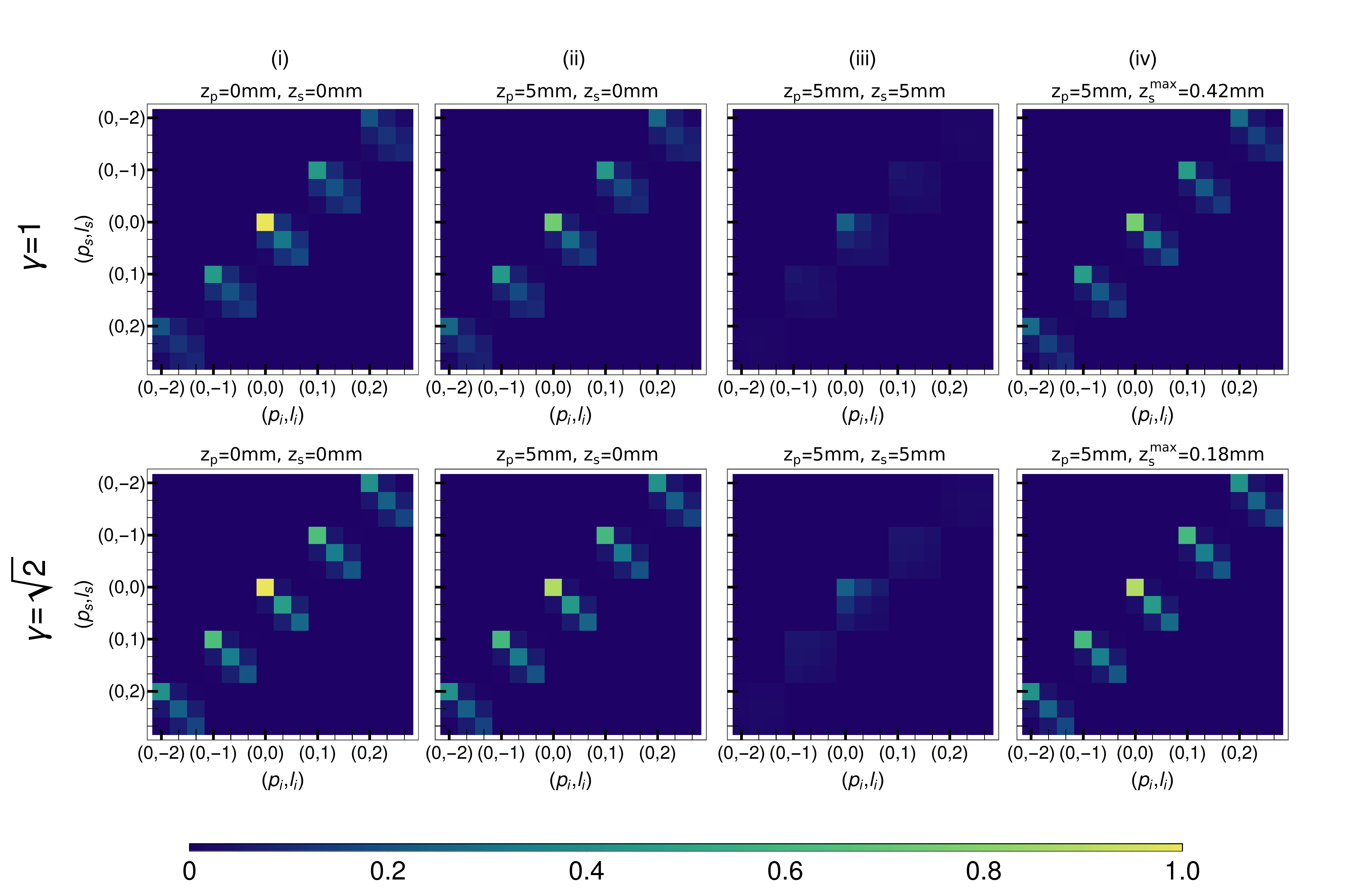}
    \caption{Same as in Fig. \ref{figure5}, but for a crystal with length $L= \SI{1}{\milli \meter}$ and different beam width ratios, $\gamma = \frac{w_p}{w_{s,i}} = 1$ (upper row) and $\gamma =\sqrt{2}$ (lower row). When all three beam focal planes are positioned in the middle of the crystal, the coincidence amplitude for $(p_s|\ell_s) = (0|0)$ and $(p_i|\ell_i) = (0|0)$ is the highest. The amplitude decreases if the focal plane of the pump beam is shifted. For this particular pump shift exists a certain shift for signal and idler $z_s=z_i \neq \SI{0}{\milli \meter}$ (d), that maximizes the amplitude and improves the efficiency. The focal plane positions of signal and idler in c) are distant from the maximizing focal plane shift. The probability of measuring photon pairs in these modes is very low.}
    \label{figure6}
\end{figure*}

The subsequent sections address how the pump, signal, and idler focal plane positions influence the spatio-temporal biphoton state. As we discussed in the previous section, we can set $z_s=z_i$ for equal signal and idler beam widths. This assumption provides for our results 99.99 \% agreement compared to the actual maximizing focal plane shits $z_s$ and $z_i$. We distinguish four different combinations of $z_p$ relative to $z_s=z_i$:
\begin{enumerate}[(i)]
\itemsep0em 
    \item The focal plane of the pump, signal, and idler are laying all in the middle of the crystal, i.e. $z_p = z_s =z_i = \SI{0}{\milli \meter}$.
    
    \item The focal plane of the pump is shifted by a certain amount $z_p$ (experimentally perhaps unintentionally and therefore not noticed). However, signal and idler beams are still positioned at the crystal center, $z_s=z_i=\SI{0}{\milli \meter}$. 
    
    \item The pump, signal, and idler focal planes are shifted by the same amount as the pump in (ii) so that they are focused on the same spot, i.e. $z_p=z_s=z_i \neq \SI{0}{\milli \meter}$.
   
    \item The pump beam is positioned on the same spot as in (ii) and (iii), but the focal plane positions of signal and idler are chosen in such a configuration, that maximizes the amplitude for the Fundamental Gaussian Mode $|C_{0,0}^{0,0}|^2$ for the given $z_p$.
\end{enumerate}

In the following, we will probe a ppKTP crystal pumped with a coherent laser operating at $\lambda_p= \SI{405}{\nano \meter}$. The procedure (i)-(iv) will be accomplished for different crystal lengths $L$ and beam width ratios $\gamma$.

\subsubsection{Influence of $z_p, z_s$, and $z_i$ on spatial biphotons state}
\label{Influence of focal plane shifts on spatial spectrum}

 \begin{figure*}[t]
\centering
    \includegraphics[width=0.95\textwidth]{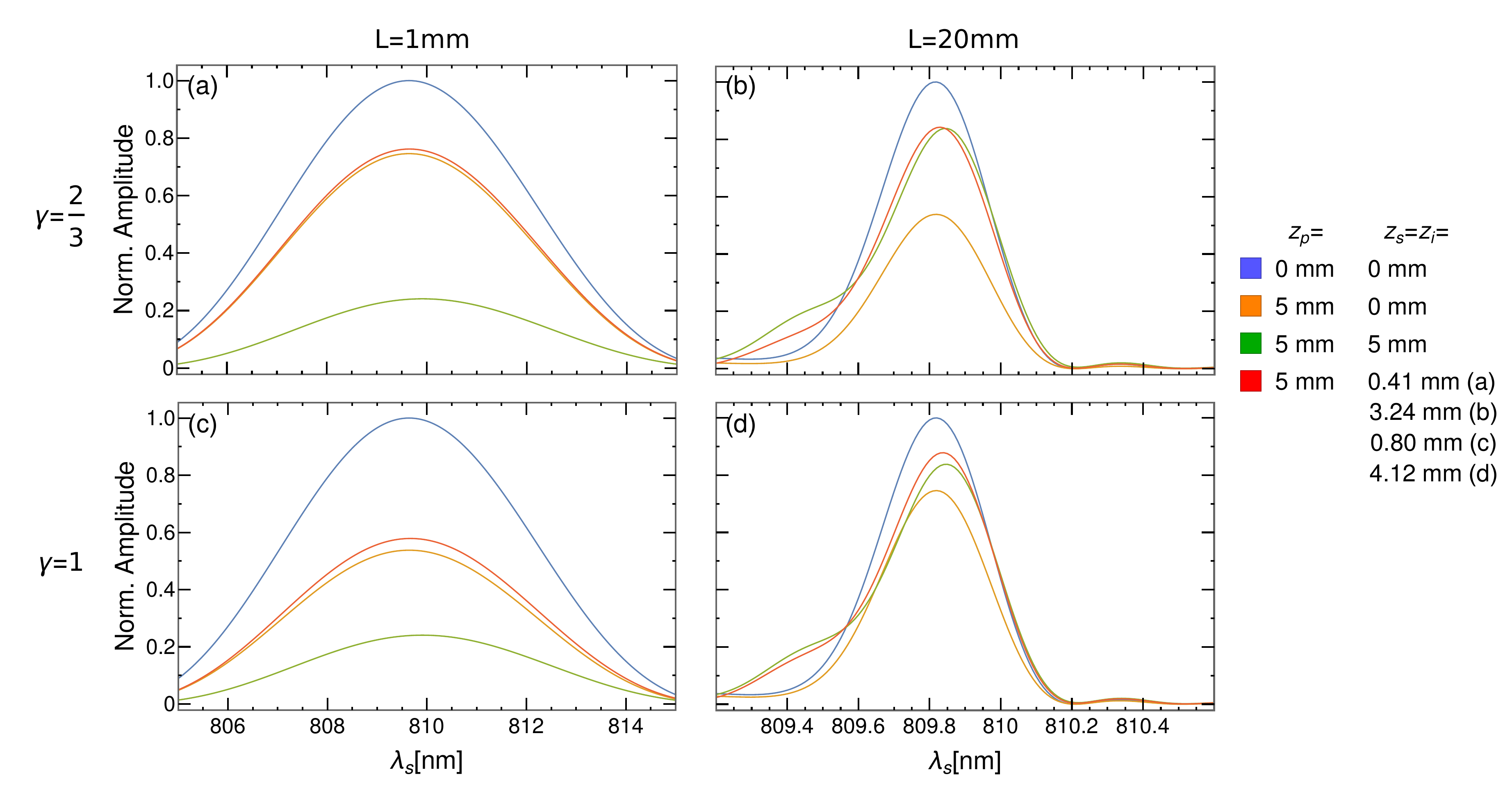}
\caption{Influence of focal plane shifts on the wavelength spectrum for signal photons. We show crystal configurations with $L= \SI{1}{\milli \meter},\SI{20}{\milli \meter}$ and $\gamma = \frac{2}{3}, 1$. The different focal plane positions of pump, signal, and idler are illustrated in each plot with different colors. Again, a pump shift of $z_p= \SI{5}{\milli \meter}$ was chosen. The red curves represent $z_s=z_i=z_s^{\text{max}}$. The highest spectral brightness is achieved when the focal planes of pump, signal, and idler are laying in the center of the crystal. Note that for thinner crystals the spectrum is much wider. It is also easy to notice that for longer crystals, focal plane shifts have a more vivid influence on the brightness. The larger the focal plane shift, the further the maximum from the initial position is shifted.}
\label{figure7}
\end{figure*}

Fistly we analyze if the spatial DOF of generated photons is affected by the change of the focal plane positions.  We apply the narrowband regime and assume signal photons filtered at $\lambda_s=\SI{810 }{\nano \meter}$. The normalized coupling efficiency of finding a pair of photons in the Laguerre-Gaussian modes with radial number $p$ and OAM number $\ell$ are shown in FIGs. \ref{figure5} and \ref{figure6}. We truncate the infinite space of mode numbers $p$ and $\ell$ to the subspace of $p_{s,i} \in \{0,1,2\}$ and $\ell_{s,i} \in \{-2, -1, 0, 1, 2\}$, where the highest contributions of the overlap amplitudes $C_{p_s,p_i}^{\ell_s,\ell_i}$ occur. The OAM conservation is easy to see for both figures since only the modes fulfilling the condition $\ell_p=\ell_s+\ell_i=0$ are non-zero. The scenarios (i)-(iv) are depicted in four columns, shown from left to right.

Fig. \ref{figure5} corresponds to the mode distribution for the crystal lengths $L= \SI{10}{\milli \meter}$ and $L=\SI{25}{\milli \meter}$, where the beam width ratio is fixed to the value $\gamma = \sqrt{2}$. The pump focal plane shift according to (ii) is $z_p=\SI{5 }{\milli \meter}$. The FGMs $(p_s|\ell_s) =(p_i|\ell_i) = (0|0)$ have always the largest amplitude in each frame. Moreover, this amplitude is the highest for the crystal with $L= \SI{10}{\milli \meter}$ when all beams are center-focused, see case (i). The probability decreases when $z_p \neq \SI{0}{\milli \meter}$ (ii). However, this can be compensated by focal plane shifts of signal and idler. If the signal and idler modes are shifted to the same position as the pump according to scenario (iii), the amplitude decreases further. The focal plane position $z_s=z_i=\SI{2.17}{\milli \meter}$ in (iv) maximizes the FGMs for $z_p=\SI{5}{\milli \meter}$. The ratios of the amplitude in comparison to (i) are in (ii) 0.89, in (iii) 0.76, and in (iv) 0.94. The more distant the focal plane positions are from $z_s^{\text{max}}$, the lower the efficiency.

The situation is different for the crystal length $L = \SI{25}{\milli \meter}$ in the second row. The scenario (i) $z_p=z_s=z_i=\SI{0}{\milli \meter}$ is no longer the best choice for achieving the highest efficiency. The ratios relative to scenario (i) now are the following: (ii) 0.89, (iii) 1.03, and (iv) 1.04. The coupling efficiency is not maximized at $z_p=z_i=\SI{0}{\milli \meter}$, since we filtered the signal at $\SI{810}{\nano \meter}$. We will show in the next section that the coupling into SMF strongly depends on the considered wavelength. Particularly, we find the optimal value at  $z_s=z_i=\SI{5.67}{\milli \meter}$ that maximize the coupling efficiency for given $z_p= \SI{5}{\milli \meter}$. 

Fig. \ref{figure6} analyzes the same as Fig. \ref{figure5}, but for for different beam width ratio values $\gamma = 1, \sqrt{2}$. The crystal length remains constant at $L=\SI{1}{\milli \meter}$. Each column indicates the scenarios (i)-(iv) of focal plane positions exactly like those described before. We make similar observations from left to right as in the upper row of Fig. \ref{figure5}: when all focal planes are in the center, the highest amplitude is achieved. These probabilities decrease again for a shifted pump. The optimal shift $z_{s,i}$ for a displaced pump focal plane ($z_p = \SI{5 }{\milli \meter}$) is close to the middle of the crystal with $z_s=z_i=\SI{0.17 }{\milli \meter}$. The efficiency is significantly reduced for the shifts $z_p = z_s =z_i = \SI{5}{\milli \meter}$, so these focal plane positions are not recommended. Higher beam width ratios $\gamma$ even allow more mode combinations, which corresponds to an increasing spiral bandwidth \cite{Yao}.

We can conclude, that the single-mode coupling efficiency of the spatial spectrum is affected by focal plane position. There is always a certain combination of $z_p, z_s$, and $z_i$ that maximizes the efficiency for a given setup. However, the positioning of all beam focal planes in the center of the crystal may not be the most effective choice to achieve the peak amplitude for all frequencies. The optimal choice of $z_p, z_s$, and $z_i$ depends strongly on the filtered frequency.

\subsubsection{Influence of $z_p, z_s$, and $z_i$ on spectral biphotons state}
\label{Influence of focal plane shits on the temporal spectrum}

\begin{figure*}
\centering
    \includegraphics[width= 0.99\textwidth]{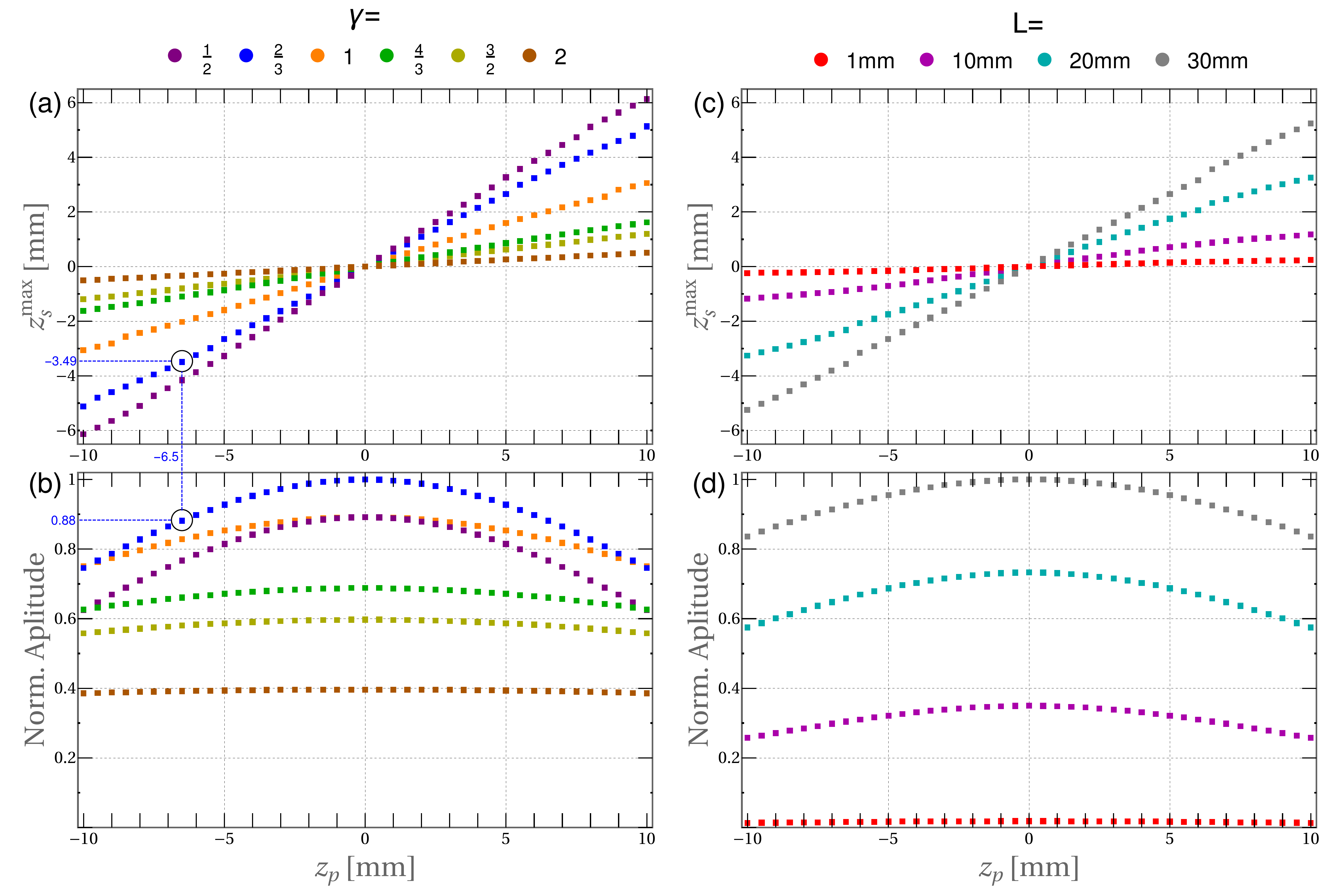}
\caption{(a) The maximizing signal/idler focal shifts $z_s^{\text{max}}$ for pump focal shifts $z_p$ from \SI{-10}{\milli \meter} to \SI{10}{\milli \meter}. We compare six different beam width ratios $\gamma$ from $\frac{1}{2}$ to $2$. The larger $\gamma$, the more horizontal the curves. A horizontal curve indicates that signal and idler focal plane should be positioned in the crystal center for all shifts of the pump focal plane. (b) For every point $z_s^{\text{max}}(z_p)$ from above in (a) the corresponding amplitude is shown. The values are normalized to the maximum amplitude. The amplitudes are comparatively small for larger beam width ratios $\gamma$, but a shift of the pump focal plane has only a small influence on the amplitude. (c) Like in a) the dependence $z_s^{\text{max}}(z_p)$ is shown. Four different crystal lengths from $\SI{1}{\milli \meter}$ to $\SI{30}{\milli \meter} $ are compared. The smaller $L$, the more horizontal the curves. (d) The corresponding amplitude for every point $z_s^{\text{max}}(z_p)$ from above in (c) is shown. The amplitudes for larger $L$ are higher, which is only possible when the demanding PM conditions in longer crystals are properly fulfilled.}
\label{figure8}
\end{figure*}

Apart from the spatial DOF, we should also analyze the influence of the focal plane shifts on the spectral DOF of the biphoton state. We consider here the spectral response of the Fundamental Gaussian Mode $ |C_{0,0}^{0,0}(\Omega)|^2$. It is enough to look only at the spectral response of the signal mode since the spectrum of signal and idler modes are symmetric with respect to the central frequency due to the continuous-wave pump, $\Omega_s=-\Omega_i$.

Fig. \ref{figure7} shows $ |C_{0,0}^{0,0}(\Omega)|^2$ for different focal plane positions and for different combinations of parameters $L = \SI{1}{\milli \meter}, \SI{20}{\milli \meter}$ and $\gamma = \frac{2}{3}, 1$. The four colors in each plot represent four different setups of combinations of $z_p = \SI{5}{\milli \meter}$ and $z_s=z_i$ according to scenarios (i)-(iv). The spectrum of signal photons is much broader for the thin crystal regime on the left compared to a thick crystal shown in the right column of Fig. \ref{figure7}. In terms of focal plane shifts, the blue curves corresponding to $z_p=z_s=z_i=\SI{0}{\milli \meter}$ show always the highest brightness. When the pump focal plane is shifted (green, red, yellow curves), the magnitude of the corresponding amplitudes changes.

Furthermore, in (d), we readily discernible that shifts of signal and idler directly shape the frequency spectrum and position of the maximum. The larger the signal and idler shifts are, the more the maximum is moved away from the value of the blue curve. We observe in Figs. \ref{figure7} (b) and (d), that for small wavelengths the blue curve lies under the green and red curves. This implies that the focusing of all modes in the middle of the crystal is not preferable anymore at this frequency. However, when considering the possible highest brightness, the condition $z_p=z_s=z_i=\SI{0}{\milli \meter}$ always provides the highest brightness.

\subsection{Generalizing the impact of $z_p$, $z_s$ and $z_i$ on spectral brightness}
\label{Response to shifted pump focal planes}

We observed from FIGs. \ref{figure5}-\ref{figure7} that for a given shift $z_p= \SI{5}{\milli \meter}$, a $z_s^{\text{max}}$ exists, which maximizes the coupling efficiency. In this section, we generalize our results and consider the dependence $z_s^{\text{max}}(z_p)$ for different beam width ratios $\gamma$ and crystal lengths $L$.  The upper plots of Fig. \ref{figure8} show the shift $z_s=z_i$ for a given $z_p$ , that maximizes the spectral brightness. The normalized amplitude is shown at the bottom of the figures. This means points that overlap vertically belong to the same $z_p$ value, see the example in Fig. \ref{figure8} (a) and (b). Note that the maximum spectral brightness is not always achieved at the same frequency for different focal plane considerations. 

In Fig. \ref{figure8} (a) we plot the maximizing signal and idler focal plane shifts $z_s^{\text{max}}$ for given pump focal shifts in the range from $\SI{-10}{\milli \meter}$ to $\SI{10}{\milli \meter}$ for a crystal of length ${L=  \SI{20}{\milli \meter}}$. Six different values for the beam width ratio are displayed. The maximizing shift $z_s^{\text{max}}$ changes linearly with $z_p$, whereby the slope of the line depends on the beam width ratio. The higher $\gamma$, the smaller the slope of the lines. Hence signal and idler focal plane can be positioned in the middle at $z_s= \SI{0}{\milli \meter}$, regardless of $z_p$. The shift of the pump focal plane has no significant impact. 

The spectral brightness is highest for $z_p=z_s=\SI{0}{\milli \meter}$ for all $\gamma$ in  Fig. \ref{figure8} (b). Whenever the pump focal plane is shifted away from the crystal center, the amplitude drops. At high values for $\gamma$, the $z_s^{\text{max}}(z_p)$ dependence becomes almost constant. The explanation is that a high beam width ratio results in a less divergent pump with big width and a more constant beam radius over the length of the crystal. This means no major noticeable change in the system and therefore less impact on the yield. The optimum beam width ratio for maximum spectral brightness in Fig. \ref{figure8} is $\gamma = \frac{3}{2}$.

It is important to mention, that the photons for every $z_s^{\text{max}}(z_p)$ value are spectrally filtered at the corresponding maximum. The frequency that maximizes the spectral brightness lies in a very small range between $\SI{809.88 }{\nano \meter}$ and $\SI{809.90 }{\nano \meter}$.

Similarly, Fig. \ref{figure8} (c) and (d) show the maximizing beam shifts for given pump shifts between $\SI{-10}{\milli \meter}$ to $\SI{10}{\milli \meter}$ at a constant beam width ratio $\gamma = \frac{3}{2} = \frac{\SI{30}{\micro \meter}}{\SI{20}{\micro \meter}}$ displaying four different crystal lengths.  Again, linear lines can be seen in the top chart. The thinner $L$, the more horizontal these lines are. This corresponds to the expectations of a thin crystal since the pump beam radius does not change significantly over the length of the crystal \cite{BaghiSteinlechnerFritzsche, Ramirez-AlarconCruz-RamirezU'Ren}. As a consequence, pump shifts have almost no effect on the spectral brightness in thin crystals. Pump focal plane shifts should be taken into account by a proper signal and idler focal plane positions $z_s^{\text{max}}$ in thicker crystals.  

\subsection{Spatio-temporal correlations between signal and idler photons}

In general, we can distinguish two kinds of correlations in the SPDC process: the correlation between signal and idler photons (see FIGs. \ref{figure5} and \ref{figure6}) or the correlation between spatial and spectral DOF of generated photons. The spatio-spectral correlation implies that the spatial characteristics of signal (idler) photons can not be considered independently of the spectral DOF, they are coupled. Mathematically, it means that the biphoton mode function can not be written as the product state of spatial and spectral DOFs  $\Phi(\bm{q}_s,\bm{q}_i,\Omega_s,\Omega_i) \neq \Phi_{\bm{q}}(\bm{q}_s,\bm{q}_i) \Phi_{\Omega}(\Omega_s,\Omega_i) $. Correspondingly, the correlation between signal and idler photons implies that the biphoton mode function can not be written as $\Phi(\bm{q}_s,\bm{q}_i,\Omega_s,\Omega_i) \neq \Phi_s(\bm{q}_s,\Omega_s) \Phi_i(\bm{q}_i,\Omega_i)$.

\begin{figure}
    \centering
    \includegraphics[width=.40 \textwidth]{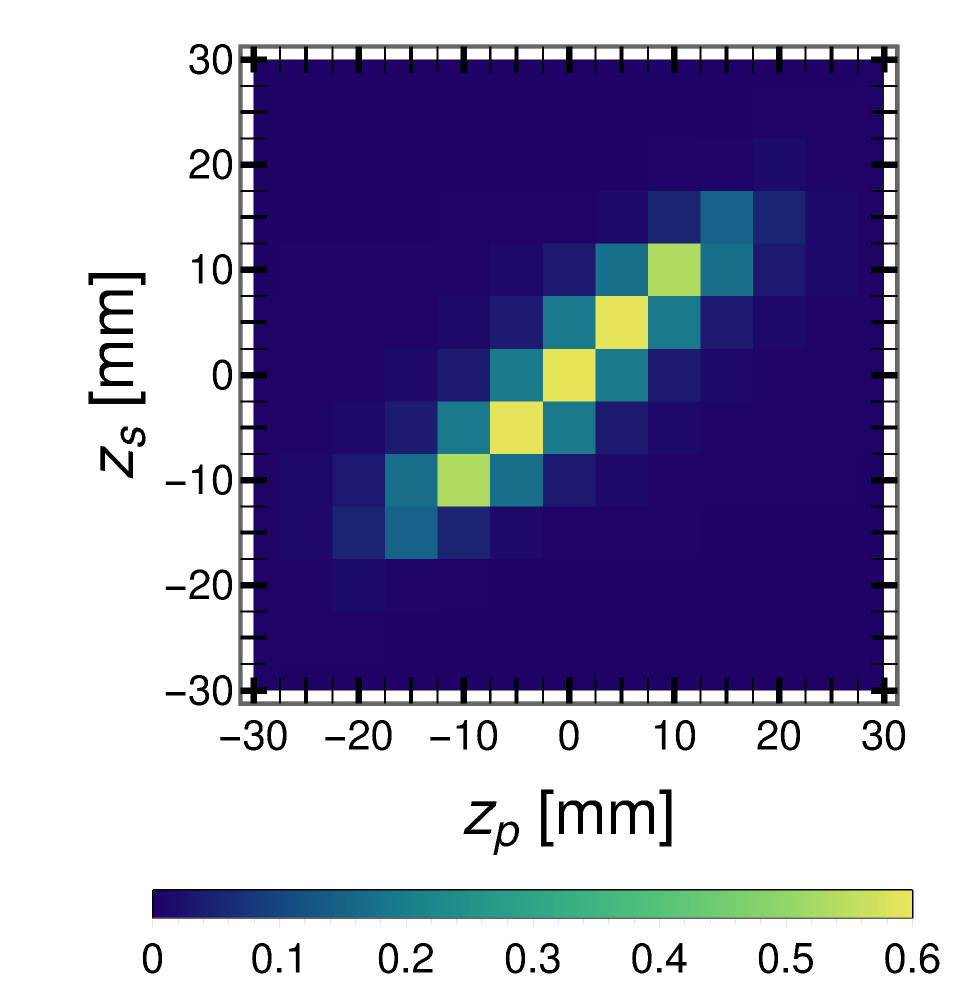}
    \caption{Spectral purity $\text{Tr}(\rho^2_{s\text{,SMF}})$ as a function of the focal plane position of the pump $z_p$ and signal beam $z_s$ with the assumption $z_s=z_i$ and a setup with crystal length $L=\SI{30}{\milli \meter}$, beam width ratio $\gamma= \frac{1}{\sqrt{2}}$ and pulse duration $T_0 = \SI{0.5}{\pico \second}$. The purity reaches its maximum when all beams are focused in the crystal center, $z_p=z_s=z_i=\SI{0}{\milli \meter}$.}
    \label{figure9}
\end{figure}

Here, we quantify explicitly how the focal plane shifts affect both types of correlations. We consider for this section the Gaussian envelope Eq. \eqref{gaussianenvelope} for the spectral DOF of the pump.

We look at the purity of the spatial (spectral) biphoton state \cite{Osorio_2008} 
\begin{align*}
    \text{Tr}(\rho^2_{\bm{q}}) &= \int d\bm{q}_s \: d\Omega_s \: d\bm{q}_i \: d\Omega_i \: d\bm{q}_s' \: d\Omega_s' \: d\bm{q}_i' \: d\Omega_s' \nonumber \\
    & \times \Phi(\bm{q}_s,\bm{q}_i,\Omega_s,\Omega_i) \: \Phi^*(\bm{q}_s',\bm{q}_i',\Omega_s,\Omega_i) \nonumber \\
    & \times \Phi(\bm{q}_s',\bm{q}_i',\Omega_s',\Omega_i') \: \Phi^*(\bm{q}_s,\bm{q}_i,\Omega_s',\Omega_i')
\end{align*}
as a measure for the correlations between space and frequency DOF. It turns our that all dependencies of the spatial purity $\text{Tr}(\rho^2_{\bm{q}})$ on $z_p, z_s$, and $z_i$ cancel out. Since $z_s$ and $z_i$ are parameters associated with detection mechanisms, this seems logical. Therefore, the spatio-spectral correlation can not be manipulated by the beam shifts $z_p, z_s$, and $z_i$. 

The situation is different for the purity of the signal (idler) photon \cite{Osorio_2008}
\begin{align}
    \text{Tr}(\rho^2_{\text{s}}) &= \int d\bm{q}_s \: d\Omega_s \: d\bm{q}_i \: d\Omega_i \: d\bm{q}_s' \: d\Omega_s' \: d\bm{q}_i' \: d\Omega_s' \nonumber  \\
    & \times \Phi(\bm{q}_s,\Omega_s,\bm{q}_i,\Omega_i) \: \Phi^*(\bm{q}_s',\Omega_s',\bm{q}_i,\Omega_i) \nonumber \\
    & \times \Phi(\bm{q}_s',\Omega_s',\bm{q}_i',\Omega_i') \: \Phi^*(\bm{q}_s,\Omega_s,\bm{q}_i',\Omega_i'),
    \label{purtiy_gen}
\end{align}
where its dependence on $z_p, z_s$, and $z_i$ does not drop off. The purity \eqref{purtiy_gen} gives the strength of the correlation between signal and idler photons, i.e., how entangled the two photons are. In the last years, one of the most important goals of photonic quantum technologies has been the reduction of the correlation between signal and idler photons. The heralded pure single photons from SPDC are believed to be a good candidate for an indistinguishable single-photon source \cite{Tambasco:16, PhysRevA.93.013801,Graffitti_2017}, which is required for a successful photonic boson sampling \cite{doi:10.1126/science.abe8770}. Usually, the spatial DOF of the biphoton state is doped off by just collecting the photons into SMF, which accepts only the Gaussian mode. We can then talk about the spectral purity of the biphoton state which can be estimated by 
\begin{align*}
    \text{Tr}(\rho^2_{\text{s,SMF}}) &= \int \: d\Omega_s  \: d\Omega_i  \: d\Omega_s' \: d\Omega_i' \:  \nonumber \\
    & \times C^{0,0}_{0,0}(\Omega_s,\Omega_i) \: C^{0,0}_{0,0}(\Omega_s',\Omega_i')\: \nonumber \\
    & \times  [C^{0,0}_{0,0}(\Omega_s',\Omega_i)]^* \: [C^{0,0}_{0,0}(\Omega_s,\Omega_i')]^*
    \label{purtiy_smf}.
\end{align*}
Fig. \ref{figure9} shows the dependence of the spectral purity $\text{Tr}(\rho^2_{s\text{,SMF}})$ on $z_p$ and $z_s$ for a crystal length $L=\SI{30}{\milli \meter}$ and pulse duration of $T_0=\SI{0.5}{\pico \second}$. The combination $z_p=z_s=z_i=\SI{0}{\milli\meter}$, which corresponds to the crystal center, yields the maximum purity. Additionally, high levels of purity can also be observed along the diagonal $z_p=z_s=z_i$.

\section{Conclusions}

In this work, we have assumed paraxial pump and collecting signal and idler beams with defined beam widths and focal plane positions. We theoretically investigated the dependence of spatial and temporal DOFs of the biphoton state on these focal plane positions. In addition, the single-mode coupling efficiency and the spectral brightness of Fundamental Gaussian Modes were studied.  Generally, the spectral brightness reaches the maximum if all involved beams are positioned in the center of the crystal. The single-mode coupling efficiency strongly depends on the frequency: for certain narrow-band filtered frequencies, positioning all focal planes in the crystal center would not attain the highest efficiency.

Depending on the setup, small deviations of the focal plane positions from the crystal center can have a large impact on the coupling efficiency. In sense of the advanced-wave picture, pump, signal, and idler focal plane shifts must compensate each other for higher efficiency. However, equal positioning of signal and idler focal planes is sufficient in most setups. Thus we advise choosing a suitable signal and idler focal plane position $z_s=z_i=z_s^{\text{max}}$ for higher efficiency if the pump beam is not center-focused. Regardless of $z_p$, $z_s=z_i= \SI{0}{\milli \meter}$ can be assumed for high beam width ratios $\gamma$ or short lengths $L$, see results in section \ref{Response to shifted pump focal planes}. 

We also find that correlations between space and frequency degrees of freedom are not affected by focal plane shifts. In contrast, the entanglement between signal and idler photons depends on the focal plane positions of involved beams. We recommend placing all focal planes of SPDC beams in the center of the crystal to achieve the highest purity.

\section{Acknowledgment}
The authors thank Carlos Sevilla-Gutiérrez for very helpful discussions and suggestions.

\section{Author contributions}
RB and BB conceptualized the original idea and developed the theoretical framework. RB conducted both, analytical and numerical computations, with assistance from BB.  SF was involved in monitoring the outcomes of this work and provided guidance throughout the research process. RB wrote the first draft of the manuscript. All authors discussed the results and contributed to the final version of the manuscript. \\

\noindent \textbf{Conflicts of Interest:} The authors declare no conflict of interest. \\
\noindent \textbf{Data Availability Statement} This manuscript has no associated data or the data will not be deposited.

\bibliography{sn-bibliography}

\end{document}